# An appraisal of Understanding Pressure Effects on Structural, Optical, and Magnetic Properties of $CsMnF_4$ and Other $3d^n$ Compounds

Fernando Rodríguez

MALTA Consolider Team, Department of Earth Sciences and Condensed Matter Physics (DCITIMAC), Facultad de Ciencias, University of Cantabria, 39005 Santander, SPAIN

rodriguf@unican.es

A recently published article in *Inorganic Chemistry*[1] offers theoretical calculations on the effects of pressure on the structural, optical, and magnetic behavior of $CsMnF_4$. Although a thorough theoretical and experimental understanding of this material is certainly warranted, there are previously published experimental results on $CsMnF_4$, as well as on other $Mn^{3+}$ fluorides, under pressure that have not been considered in the aforementioned article.[1] These findings raise questions about the accuracy of the theoretical estimates and the validity of the methodology used to determine the reported structural, optical, and magnetic properties. This paper intends to provide the readers of *Inorganic Chemistry* and the broader scientific community with a perspective on a publication[1] that directly contradicts interpretations presented in prior research.

The main scientific concerns are summarized in the following six points.

**1) Structure of $CsMnF_4$ at high pressure**

The authors theoretically describe the structural evolution of $CsMnF_4$ in terms of a tetragonal structure *P4/n* (0-40 GPa), which experiences a structural phase transition at 37.5 GPa to another tetragonal phase (*P4*) that is stable up to at least 50 GPa. In this structural evolution, the ground state of $Mn^{3+}$ is high spin (S = 2). The point is that there are three publications[2-4] dealing with the structural evolution of $CsMnF_4$ with pressure by x-ray diffraction (XRD): energy dispersive XRD at the Daresbury Synchrotron, UK,[2,3] and angle dispersive XRD at the European Synchrotron Radiation Facility (ESRF), France.[4] These independent works concluded that the tetragonal structure *P4/n* of $CsMnF_4$ is unstable above 2 GPa, transitioning to a monoclinic structure stable up to at least 16 GPa (**Fig. 1**).[4] Energy dispersive XRD data[2,3] conclude that a tetragonal to orthorhombic phase transition at 1.4 GPa followed by a monoclinic transition above 6 GPa (see Orthorhombic to Monoclinic section of Ref. 3). None of these publications were discussed in the article in relation to the experimentally observed structural evolution – Refs. 2 and 4 were not cited– and therefore, this low-pressure structural transition calls into question all results derived from a tetragonal structure above 2 GPa. Although Ref. 3 was cited in the article,[1] it literally states that "No signs of a structural phase transition around 1.4 GPa, early suggested by Moron et al.,[21] have been found in the optical measurements on $CsMnF_4$" (page 13233 in Ref. 1). Besides the recognition of a structural phase transition around 1.4 GPa earlier reported by Moron et al.,[3] the authors ignored two other publications[2,4] not cited in Ref. 1, which confirm the tetragonal to monoclinic structural phase transition (around 2 GPa in Ref. 4). In addition to the XRD results, reference 4 also reports precise optical data in the 0-16 GPa range, which provides clear evidence of the influence of the phase transition at 2 GPa in the optical spectra of $CsMnF_4$ (**Fig. 1**).

The measured spectra show that the pressure shift rate of the $^5B_1 \rightarrow {}^5A_1$ transition energy at 1.89 eV changes from -42 meV/GPa to +2 meV/GPa at 2 GPa, coinciding with the tetragonal to monoclinic phase transition (**Fig. 1**).[4] It must be emphasized that Refs. 2-4 are the only ones dealing with the crystal structure of $CsMnF_4$ under high pressure conditions, and were not considered, or cited[2,4] in the reported work,[1] although one of the authors cited Ref. 4 in a previous publication.[5] Therefore, the structural phase transition detected by XRD by two different research groups, at two different synchrotron facilities, between 1-2 GPa was neither considered in the article[1] nor detected by using the authors' DFT methodology. Furthermore, the structural evolution of $CsMnF_4$ with increasing pressure reported in reference 4 coincides with the structural evolution in the series *A*$MnF_4$ (*A*: Cs, Rb, Tl, K, Na) with decreasing the unit cell volume[2-4b] as it can be seen in **Figs. 2 and 3**.

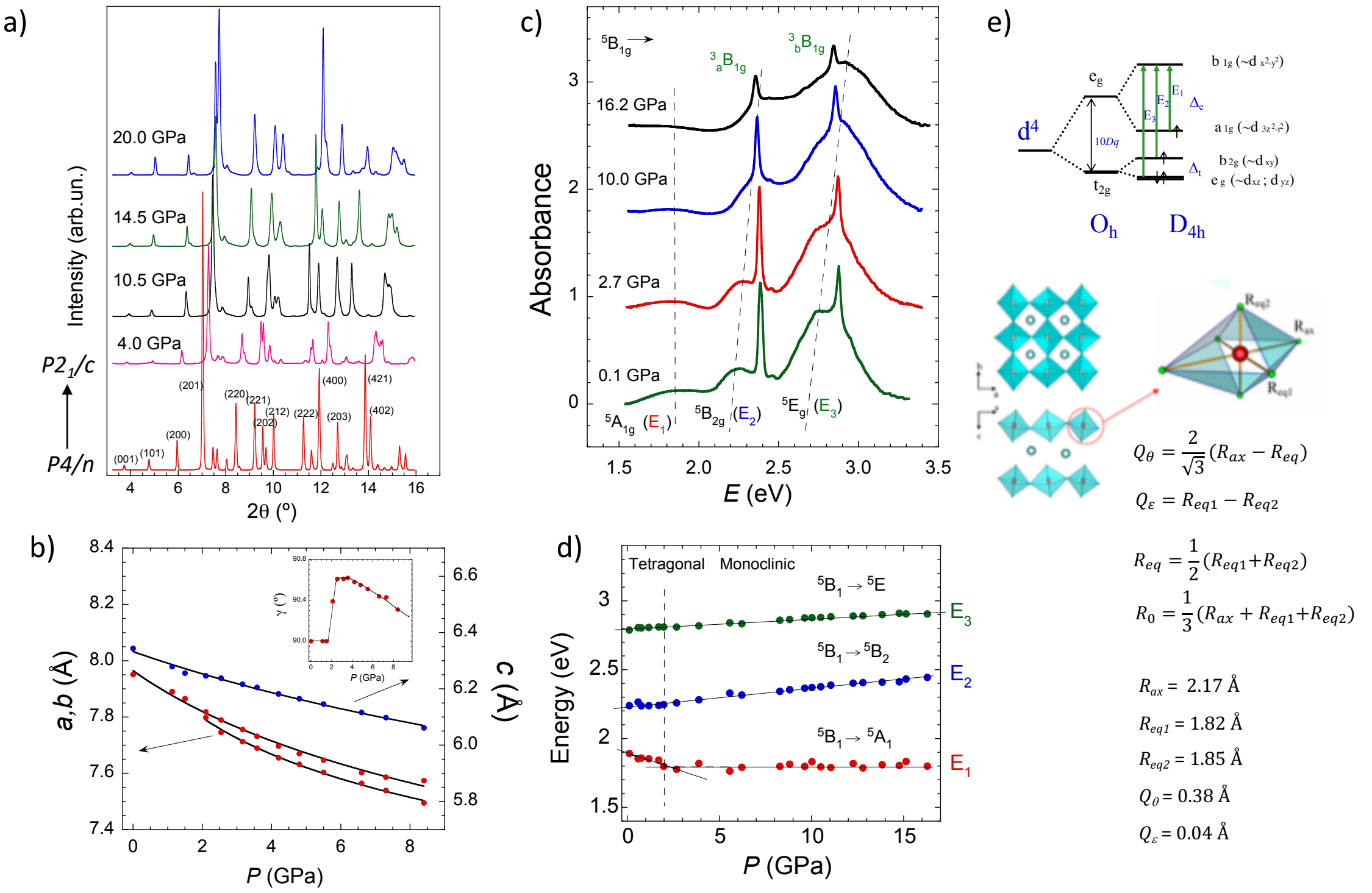

**Figure 1.** (**a**) Powder x-ray diffraction (XRD) of $CsMnF_4$ as a function of pressure, revealing a structural phase transition from tetragonal to monoclinic phase at 2 GPa. (**b**) Pressure dependence of the unit cell parameters *a*, *b*, *c* and the monoclinic angle $\gamma$ (shown in the inset), as determined from XRD data. (**c**) Evolution of the optical absorption spectrum of $CsMnF_4$ within the pressure range the 0–16 GPa. The labeled peaks correspond to electronic transitions associated with the axially elongated $D_{4h}$ symmetry of the $(MnF_6)^{3-}$ complex. The dotted lines serve as visual guides to track the pressure-induced shifts of the Jahn-Teller-related three-band structure, which originates from the Jahn-Teller effect. Observe the decrease in the intensity of the narrow peaks with increasing pressure. (**d**) Detailed pressure dependence of the transition energies $E_1$, $E_2$ and $E_3$ with pressure in $CsMnF_4$. The straight lines represent linear fits to the data, with slopes indicating the pressure shifts. The slopes are -42 meV/GPa (0-2 GPa) and +2 meV/GPa (2-16 GPa) for $E_1$, and +14 meV/GPa and +9 meV/GPa for $E_2$ and $E_3$, respectively, in the 2-16 GPa range: $\Delta_t$ = 556 - 5 *P* (meV/GPa). **e)** Schematic diagram illustrating the splitting of the of the $Mn^{3+}$ *d*-levels in octahedral ($O_h$) and elongated tetragonal ($D_{4h}$) coordination environments, showing the $E\otimes e$ Jahn-Teller effect. The bottom right panel depicts the ambient-pressure crystal structure of the layered perovskite $CsMnF_4$ (space group: *P*4/*n*), including in-layer and intralayer views, the elongated $(MnF_6)^{3-}$ complex with axial ($R_{ax}$,) and equatorial ($R_{eq1}$ and $R_{eq2}$) Mn-F bond distances, and the normal coordinates, $Q_\theta$ and $Q_\varepsilon$, which represent tetragonal and rhombic distortions, respectively. Note the antiferrodistortive structure shown by the $(MnF_6)^{3-}$ octahedra within the *a*,*b* layer.

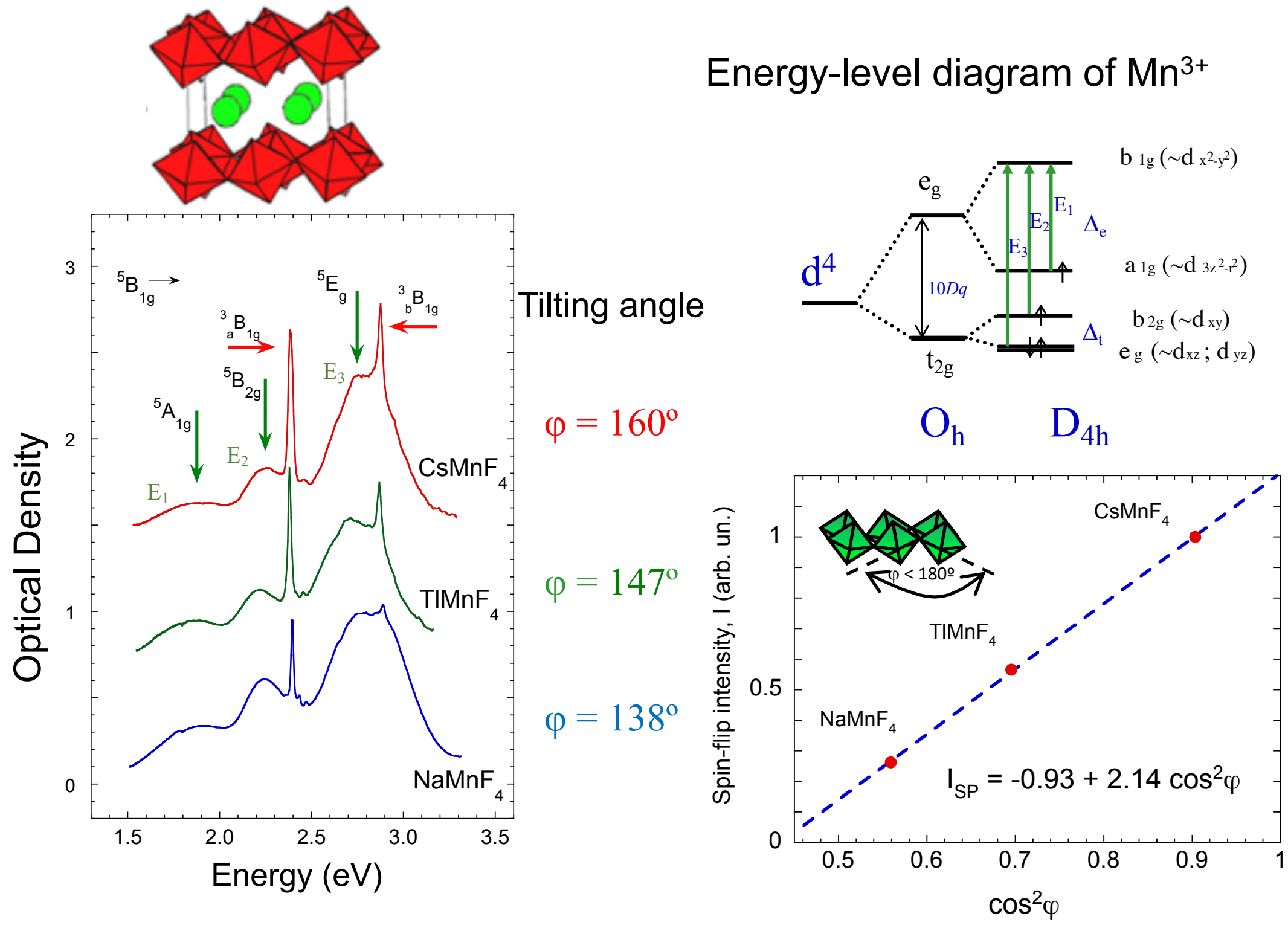

**Figure 2.** Optical absorption spectra of $NaMnF_4$, $TlMnF_4$ and $CsMnF_4$ single crystals. The Mn-F bond distances ($R_{ax}$ and averaged $R_{eq}$) are 2.15, 1.82 Å for $TlMnF_4$ and, 2.17, 1.84 Å for $NaMnF_4$ and $CsMnF_4$. The Mn-F-Mn bond angle (φ) is indicated on the right. Green (vertical) arrows represent spin-allowed crystal-field transitions ($E_1$, $E_2$, $E_3$), while red (horizontal) arrows indicate spin-flip peaks ($E_{SP1}$, $E_{SP2}$). The integrated intensity of the spin-flip peaks decreases with the tilting angle, θ (180−φ), where φ is the Mn-F-Mn bond angle. This variation shows a linear dependence with $\cos^2\theta = \cos^2\varphi$, which is proportional to the exchange constant *J*,[4b] demonstrating the exchange-induced electric-dipole mechanism for the spin-flip transitions.[6] Errors are 0.05 for relative intensity and 0.005 for $\cos^2\varphi$.

## 2) Pressure dependence of the optical and electronic properties

The pressure dependence of the optical and electronic properties derived theoretically from the tetragonal phase[1] are unable to explain the experimental results by optical spectroscopy published previously.[4,6-8] $CsMnF_4$ in the tetragonal phase is a uniaxial crystal instead of a biaxial crystal attained in the monoclinic structure (*P* > 2 GPa). Furthermore, the structure of the $Mn^{3+}$ one-electron *d*-orbitals obtained theoretically,[1] predicts that the energy of the first absorption band associated with the $^5B_1 \rightarrow {}^5A_1$ (denoted by the one-electron orbital transition $3y^2-r^2 \rightarrow x^2-z^2$ in Fig. 4 of Ref. 1), decreases continuously from 1.92 eV to 1.43 eV in the 0-40 GPa range, while experimentally it changes from 1.89 to 1.81 eV in the 0-2 GPa range (tetragonal phase) but blueshifts by only 0.03 eV in the 2-16 GPa range (monoclinic phase)[4,6] giving a total shift of +0.07 eV at 37 GPa.[4,8] It means that the band shifts from 1.89 eV at ambient pressure to 1.88 eV at 37 GPa, which is two orders of magnitude lower than the shift predicted theoretically of -0.49 eV in the same pressure range (**Fig. 1**). Neither the magnitude of the pressure band shift, nor the two opposite pressure rates below and above 2 GPa were accounted for theoretically.[1] References 4 and 7 were not cited in Ref. 1, and their experimental results could have guided the authors to search the actual structural evolution of $CsMnF_4$ as well as its associated properties. It must be noted that the pressure-induced **blueshift** of the first absorption band experimentally observed in the $CsMnF_4$ monoclinic phase above 2 GPa is also observed in $NaMnF_4$ in the 0-6 GPa range, which has a monoclinic structure at ambient conditions (**Fig. 3**).[7,8] It seems that the preservation of the tetragonal structure in the whole explored pressure range[1] is unable to explain the observed optical properties.

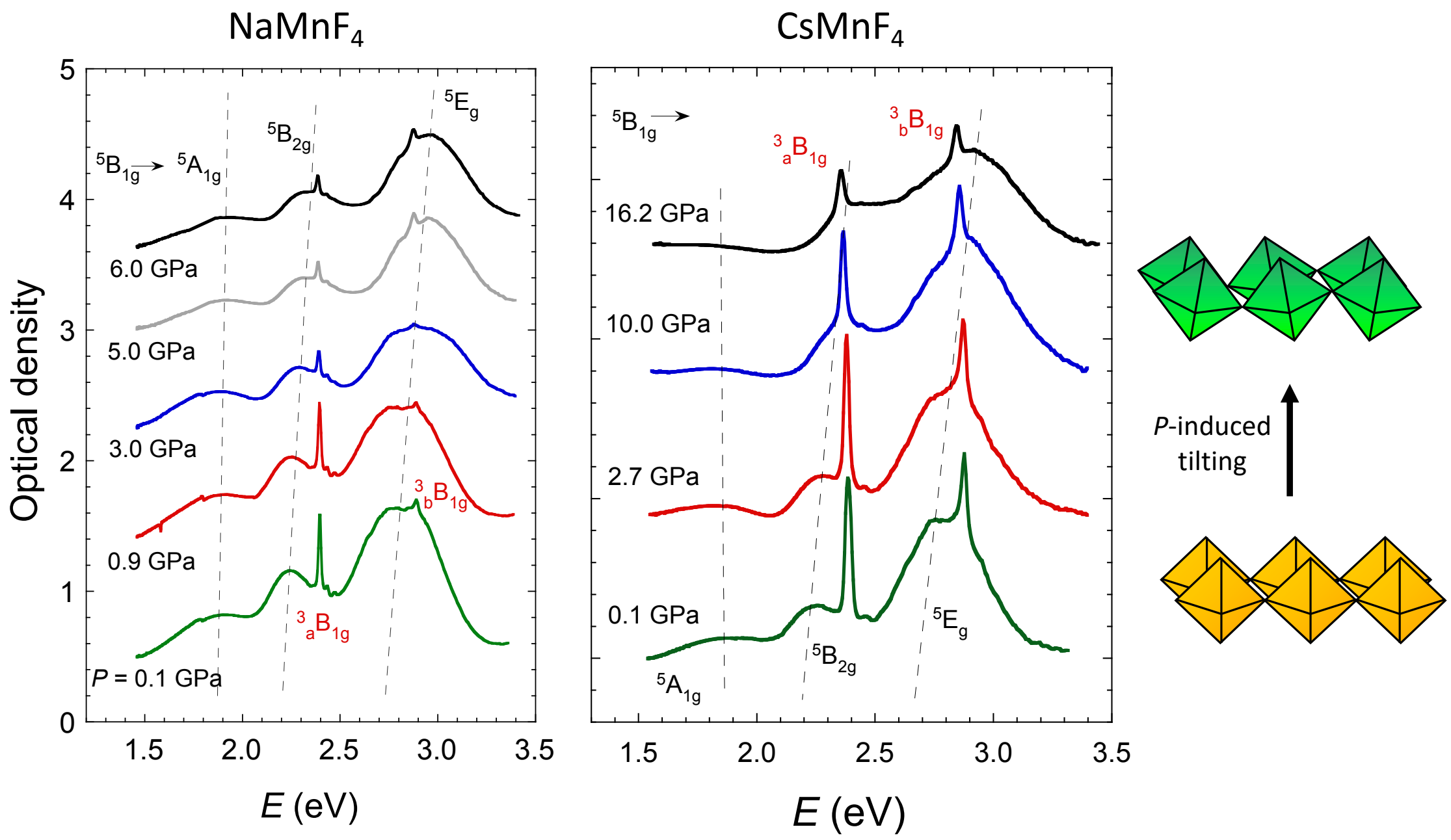

**Figure 3.** Variation of the optical absorption spectrum with pressure of $NaMnF_4$ (0-6 GPa) and $CsMnF_4$ (0-16 GPa) single crystals at room temperature (upstroke). Broken lines illustrate the pressure-induced shift for the three broadbands. Note that the spin-flip transition oscillator strengths decrease with pressure, while the absorbance of the three broad bands, which reflect the low symmetry Jahn-Teller $D_{2h}$ (nearly $D_{4h}$) splitting, varies slightly. This variation is interpreted in terms of pressure-induced $(MnF_6)^{3-}$ tilting decreasing the in-plane Mn-F-Mn exchange interaction.

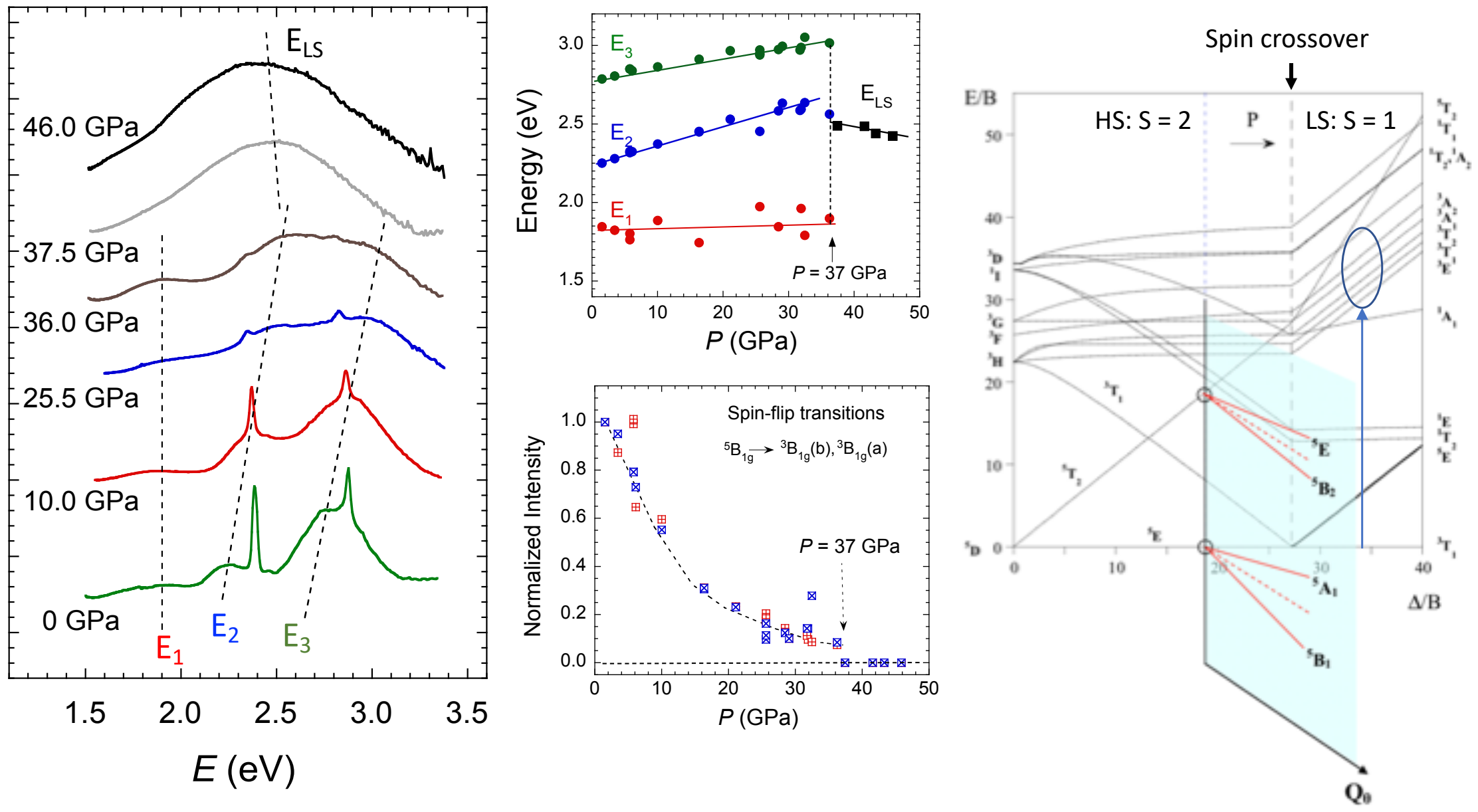

**Figure 4.** Pressure-dependent optical absorption spectrum of $CsMnF_4$ single crystal (0–46 GPa, room temperature, upstroke). Broken lines illustrate the pressure-induced shifts for the three broadbands. A sharp spectral change occurs at 37 GPa. The right panel shows the Tanabe-Sugano diagram for $Mn^{3+}$ ($3d^4$) in elongated-$D_4$ (blue shaded) and $O$ symmetries, with corresponding crystal-field excitations. Top right: pressure dependence of $E_1$, $E_2$, and $E_3$. Note that $E_2$ and $E_3$ exhibit a significant blueshift, while $E_1$ remains nearly constant (energy error: 10 meV). The decrease in intensity of spin-flip transitions $^5B_{1g} \rightarrow {}^3B_{1g}$ (a+b) at 2.380 and 2.873 eV, which vanish above 37 GPa, indicates pressure-induced tilting. Up to 36 GPa, the Jahn-Teller related broadband structure and spin-flip peaks are observed, undergoing abrupt jumps at the critical pressure $P_C$=37 GPa. Above this pressure, the spectrum becomes a structureless broadband that can be interpreted in terms of spin-allowed transitions from a low-spin (S=1) ground state $^3T_{1g}$ (within an $O_h$ coordination) to various excited states $^3\Gamma_i$ (encircled in the Tanabe-Sugano diagram). This suggests a high-spin to low-spin crossover in $CsMnF_4$ at 37 GPa. Reprinted from Reference 6. Copyright 2007 American Physical Society

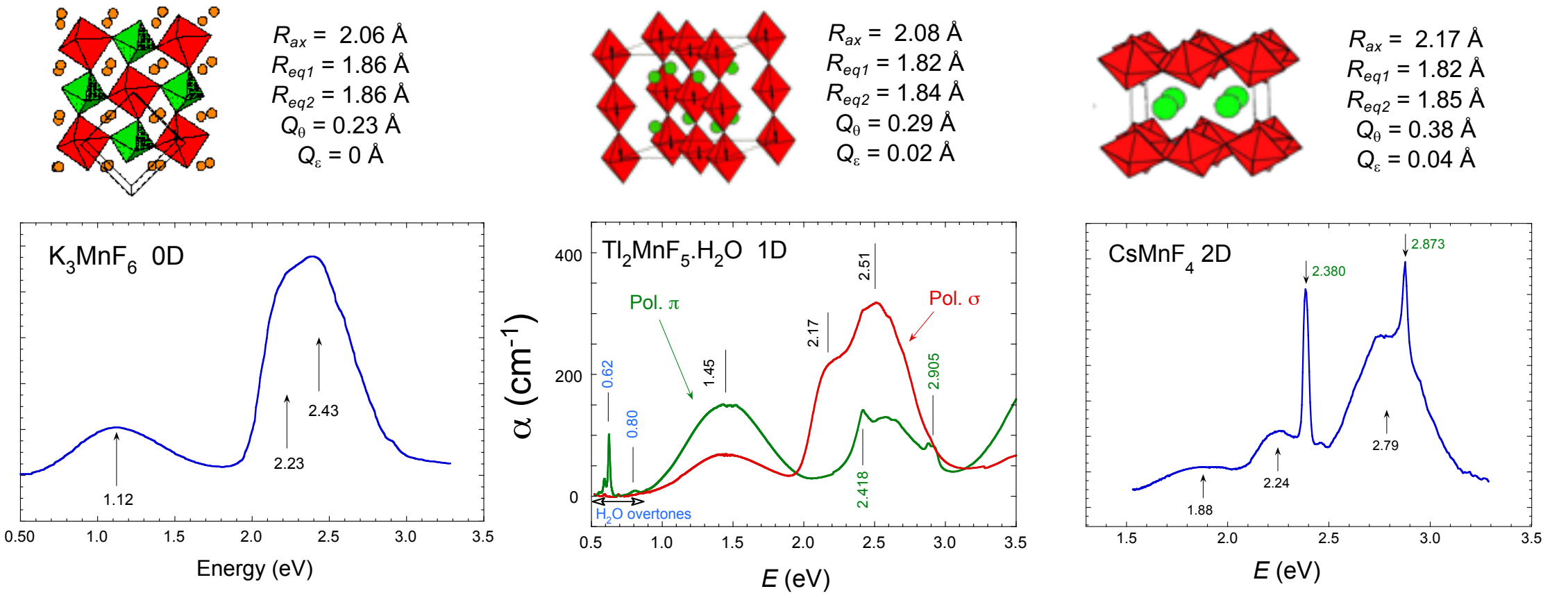

**Figure 5.** Optical absorption spectra of single-crystal $Mn^{3+}$ fluorides with different dimensionality: $K_3MnF_6$ (0D), $Tl_2MnF_5.H_2O$ (1D), and $CsMnF_4$ (2D). Top: Schematic view of the crystal structure and local structural parameters for $(MnF_6)^{3-}$. Bottom: optical absorption spectra of the three compounds measured at ambient conditions. Transition energies at the band maximum are indicated- The polarized spectra of 1D $Tl_2MnF_5.H_2O$ were measured with the light electric vector parallel (π) and perpendicular (σ) to the edge-sharing $MnF_4F_{2/2}$ chains. All spectra exhibit three broad bands (spin-allowed transitions in nearly $D_{4h}$ $(MnF_6)^{3-}$ complex, energies in eV) and two narrow peaks (2.4 and 2.9 eV, spin-flip transitions in $3d^4$). Broad band energies and associated 3d splitting depend on $D_{4h}$ distortion $Q_\theta$ (and $D_{2h}$ rhombic distortion $Q_\varepsilon$), while spin-flip peaks are less sensitive to pressure or local $Mn^{3+}$ distortion. Spin-flip peak intensity, unlike broad band intensity, strongly depends on the exchange interaction: maximum in 2D, weaker in 1D, not observed in 0D. These transitions are polarized along the Mn-F-Mn superexchange direction, consistent with selection rules for exchange-induced electric-dipole transitions.

### 3) Optical band assignment in $Mn^{3+}$ fluorides.

The band assignment of the crystal-field spectra related to *d-d* transitions in $CsMnF_4$ as well as the high-spin (HS) to low-spin (LS) transition detected by optical spectroscopy at 37 GPa reported elsewhere (**Fig. 4**),[6] the spectra of which are reproduced in Fig. 4 of Ref. 1, are questioned in the article.[1] Although no alternative interpretation to the assignment of Ref. 6 is given in the article,[1] it must be noted that the optical absorption band assignment of $Mn^{3+}$ fluorides is based on spectroscopic results reported elsewhere.[9] This work, which is not cited in Ref. 1, deals with single-crystal polarized optical absorption spectroscopy and the temperature dependence of the various bands appearing in the optical spectra. Based on their transition energy, spectral shape and oscillator strength, as well as their temperature dependence (**Figs. 5-7**), it was concluded that there are two types of bands in the optical spectra: three broad bands associated with single-electron transitions within the *d*-orbitals of $Mn^{3+}$ in an almost $D_{4h}$ local environment, and two spin-flip narrow peaks arising from the $3d^4$ electronic configuration of $Mn^{3+}$ (**Fig. 5**).

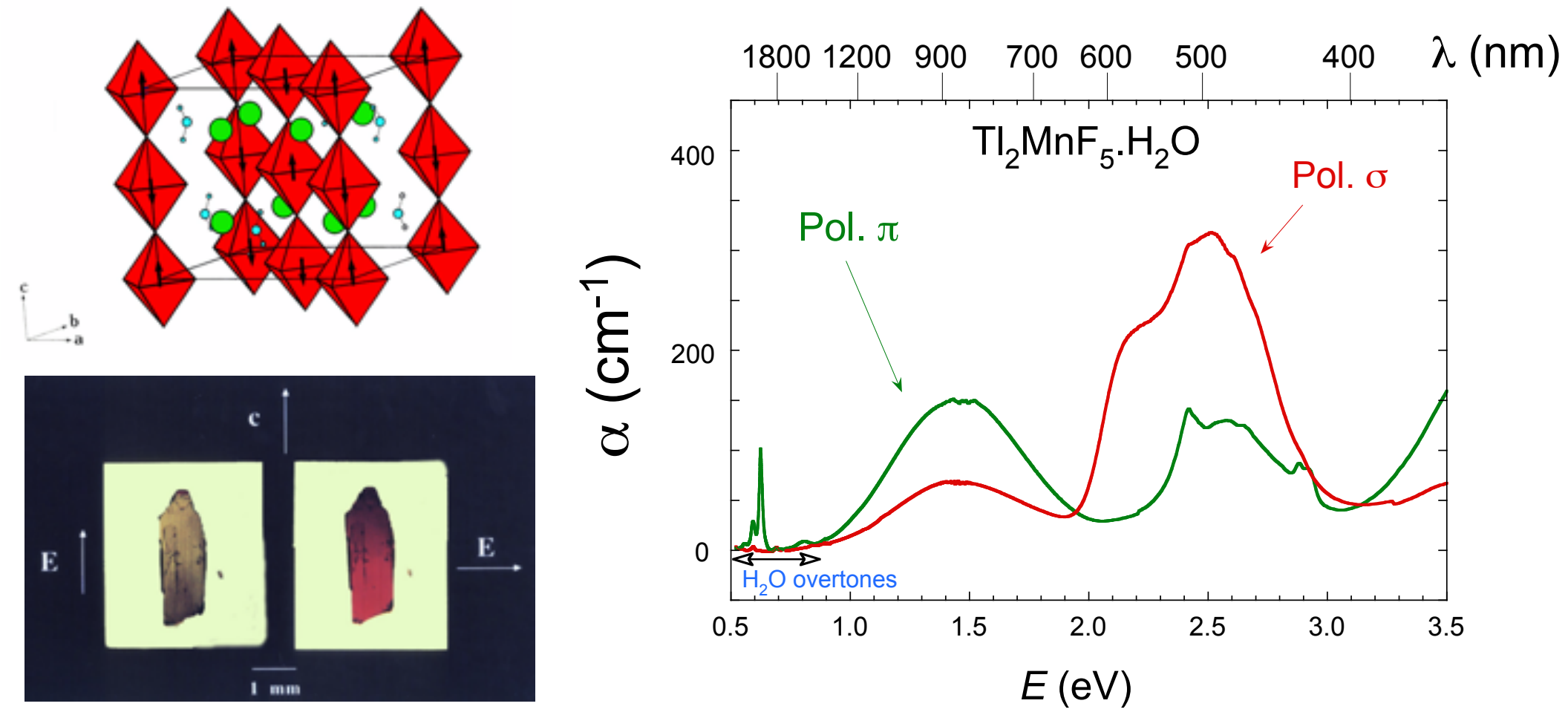

**Figure 6.**

(a) **Schematic of the 1D $Tl_2MnF_5.H_2O$ crystal's crystallographic and magnetic structures**. It is orthorhombic (*Cmcm* space group) with lattice parameters $a$ = 9.688 (2) Å , $b$ = 8.002 (1) Å and $c$ = 8.339 (1) Å at room temperature. The structure consists of linear chains of trans-connected $[MnF_4F_{2/2}]$ octahedra along the *c* axis. These octahedra exhibit near-$D_{4h}$ symmetry, elongated along the chain due to the Jahn-Teller effect and crystal anisotropy. The Mn-F-Mn bond angle (β) is 179.2˚, close to 180˚. Magnetically, the crystal is antiferromagnetic with an intrachain exchange constant (*J*) of 15 $cm^{-1}$. A 3D magnetic ordering occurs below the Néel temperature ($T_N$) of 28 K, with spins aligned parallel to the **c** -axis.[9]
(**b**) **Dichroism of a $Tl_2MnF_5.H_2O$ single-crystal**. The crystal appears olive green under polarized illumination with the light electric vector (***E***) parallel to the *c*-axis, and red with ***E*** perpendicular.
(**c**) **Polarized optical absorption spectra**. Spectra are shown for ***E*** parallel (π) and perpendicular (σ) to the chain (*c*-axis). The σ–polarized spectrum displays three broad bands ($E_1$, $E_2$ and $E_3$), similar to other $A_nMnF_m$ (n = m-3 = 1-3) compounds (**Fig. 5**). The π–polarized spectrum shows two additional narrow peaks ($E_{SP1}$, $E_{SP2}$) and two broad bands ($E_1$ and $E_3$); the $E_2$ band is absent.[9] The polarization and temperature dependence (**Fig. 7**) confirm $E_1$, $E_2$ and $E_3$ as crystal-field transitions between one-electron *d* orbitals in elongated $D_{2h}$, nearly $D_{4h}$, symmetry, and $E_{SP1}$ and $E_{SP2}$ as spin-flip transitions within the $3d^4$ configuration.
Adapted with permission from Reference 9. Copyright 1994 American Chemical Society.

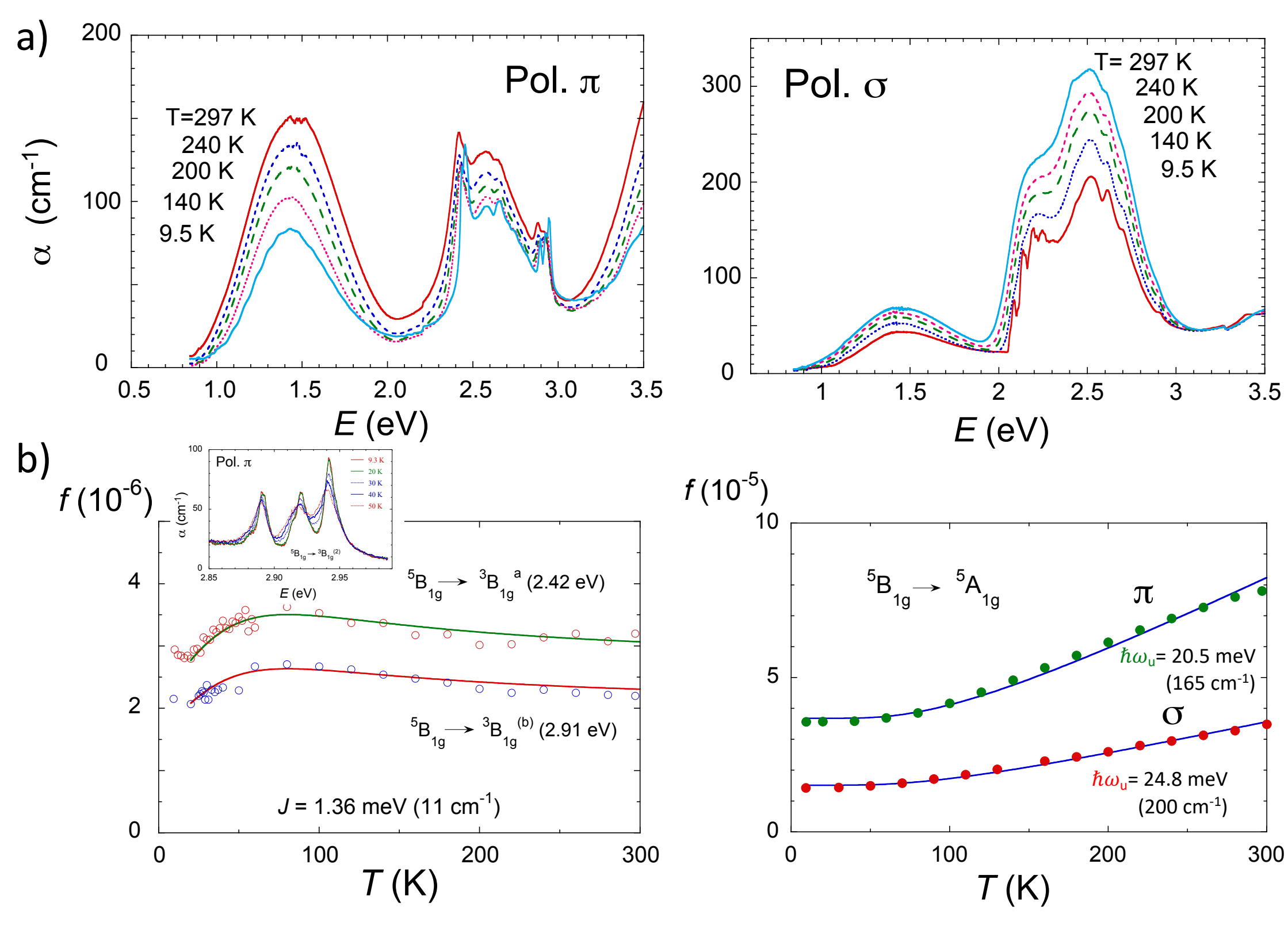

**Figure 7.**

(a) **Temperature-dependent π- and σ-polarized absorption spectra of $Tl_2MnF_5.H_2O$ (9.5-297 K).**

(**b**) **Left: Temperature dependence of the oscillator strength for the spin-flip transitions $^5B_{1g} \rightarrow {}^3B_{1g}$(a,b) in π-polarization**. Solid lines are fits to the Equation1 (exchange-induced transition mechanism)[9] with $J$ =1.36 meV, $f(0)$ = 2.0 x $10^{-6}$ ($E_{SP1}$= 2.42 eV) and 1.5 x $10^{-6}$ ($E_{SP2}$= 2.91 eV). The inset shows a magnified view of the $E_{SP2}$ temperature dependence. Note that these peaks are π-polarized and thus are not observed in σ-polarization (Figs. 6 and 7a).

**Equation 1**:

$$f(T) = \frac{2f(0)}{(2S-1)}\left[S(U^2-1)+\frac{4SU(S+1)UV}{2V}\right]$$

Where $f(0)$ is the 0 K oscillator strength, $U$=coth$V$-1/$V$, V=2$JS(S$+1)/$k_B$T , $S$ = 2 and $J$ (<0) is the intrachain exchange constant.[9]

**Right: Temperature dependence of the oscillator strength for the $^5B_{1g} \rightarrow {}^5A_{1g}$ broad band (1.45 eV) in π and σ polarizations**. Solid lines are fits to $f(T) = f(0)\,\coth(\hbar\omega_u/2k_BT)$ (vibronic mechanism, parity-forbidden electric-dipole transitions activated by odd parity vibrations $\omega_u$). The different temperature dependences and polarization reveal distinct activation mechanisms: exchange for spin-flip peaks and vibronic for spin-allowed broad bands.

It is experimentally shown that the former ones are electric-dipole spin-allowed transitions, whose oscillator strength increases with increasing temperature by coupling to odd parity vibrations within the $(MnF_6)^{3-}$ complex. The spin-flip transitions are shown to be electric-dipole spin-forbidden transitions within a single $Mn^{3+}$ ion but are activated by the spin-effective mechanism in exchange-coupled systems via the Mn-F-Mn superexchange pathway (**Fig. 7**).[9] The temperature dependence of their oscillator strength demonstrates the spin-flip nature of these narrow peaks observed in the optical spectra of exchange-coupled systems.[6-9] In addition, these spin-flip transitions can be easily identified in the optical spectra because, unlike the spin-allowed broad bands whose energies are strongly dependent on the $(MnF_6)^{3-}$ octahedral distortion, their energy is not as sensitive to the distortion (**Fig. 5**).[6-9] Spin-flip transitions are within 0.05 eV at the same position in all series of manganese (III) fluorides: $E_{SP1}$ = 2.380 eV and $E_{SP2}$ = 2.873 eV in 2D $CsMnF_4$,[6] 2.397 and 2.890 eV in 2D $NaMnF_4$,[7,8] 2.380 and 2.860 eV in 2D $TlMnF_4$ (2D fluorides)[7] and, 2.418 and 2.914 eV in the 1D $Tl_2MnF_5.H_2O$.[9]

In this respect, the *d*-orbital electronic structure derived theoretically for $CsMnF_4$ in the article,[1] does not agree with the measured spectra (**Fig. 4**),[6] which are reproduced in Fig. 4 of Ref. 1. The spectra show three broad bands associated with single-electron transitions from $3z^2$-$r^2$, xy, and the nearly degenerate (xz, yz) to $x^2$-$y^2$, in order of increasing energy (note that the local z-axis in Ref. 6 refers to the long Mn-F bond in $CsMnF_4$ of the nearly $D_{4h}$ $(MnF_6)^{3-}$ complex). In the article,[1] the xy, xz, and yz *d*-orbitals have different energies with respect to the $x^2$-$y^2$ level: 2.2, 2.5 and 2.8 eV (Fig. 7 of Ref. 1). Except for the first absorption band $3z^2\text{-}r^2 \rightarrow x^2\text{-}y^2$ at about 1.9 eV, the other three transition energies cannot give rise to a two-broad-band structure in the optical spectra as observed experimentally (**Fig. 5**), but to three almost equally spaced broad bands, which would probably give rise to a structureless broad band instead of the two well-resolved bands observed experimentally.[6] Furthermore, **the calculations performed in the article deal with one-electron *d*-orbitals and not with the states arising from the $d^4$ electronic configuration, and thus the observed spin-flip transitions are missed in the reported calculations**.[1] This is an important point as the authors should know that the spin-flip transitions observed in the spectra of $CsMnF_4$ (Refs. 4,6) are crucial to explain the spin transition at 37 GPa on the basis of optical spectroscopy (**Figs. 4, 5**).[6]

**4) Magnetic properties of $CsMnF_4$**

Regarding the magnetic properties, it is theoretically found that $CsMnF_4$ undergoes an abrupt change from ferromagnetic to antiferromagnetic at 10 GPa within the tetragonal phase.[1] However, there is a misunderstanding with the magnetic measurements under high pressure conditions reported elsewhere.[10] This paper reports a singular magnetic behavior above 2 GPa with a reduction of the Curie temperature and a sharp decrease of the saturation magnetization of $CsMnF_4$ above 3.3 GPa, which the authors explain in terms of a progressive transition from ferromagnetic to antiferromagnetic with a permanent ferromagnetic component, which the authors attribute to canting antiferromagnetism.[10] This behavior is overlooked in the article[1] and therefore, properties observed in the monoclinic phase are not predicted in the tetragonal phase. In fact, this behavior has been explained along the $AMnF_4$ series of layered perovskites, where the only tetragonal member of the series is the ferromagnetic $CsMnF_4$, the remaining compounds having a monoclinic structure are all antiferromagnetic.[2,10,11] Moreover, Refs. 2 and 3 show that the AFM Neel temperature correlates with the Mn-F-Mn tilting angle in the monoclinic phase, a figure which cannot be reproduced in the same way using a tetragonal phase due to structural constrains.

**5) High-spin to low-spin transition at 37 GPa.**

The change in the optical spectra (Fig. 4 of Ref. 1; or Fig. 2 of Ref. 6) associated with a HS$\rightarrow$LS crossover transition at 37 GPa and room temperature in the original paper[6] (**Fig. 4**) is now questioned and associated with a tetragonal to tetragonal phase transition *P4/n*$\rightarrow$*P4* at 37.5 GPa, both phases being HS (S=2). However, this interpretation lacks the spectral evolution at the crossover transition point. There are three main reasons for retaining the HS to LS transition: 1) The three broad bands observed below 30 GPa, associated with the strong octahedral distortion due mainly to the E$\otimes$e JT effect, transform into a structureless broad band characteristic of a nearly octahedral coordination. The T$\otimes$e JT effect in the LS $^3T_1$ ground state ($O_h$) is a factor of 4 weaker than the E$\otimes$e HS $^5E$ ground state strongly distorted by the JT effect;[6,12] its stabilization energy in LS is an order of magnitude smaller than in HS. 2) The spin-flip transitions $E_{SP1}$ and $E_{SP2}$, which are characteristic of a HS ground state, disappear completely above 37 GPa. According to the Tanabe-Sugano diagrams for $d^4$ ions,[12,13] these transitions are absent in a LS ground state ($^3T_1$), making them a precise spectroscopic probe for layered *A*$MnF_4$ (*A*: Cs, Na, Tl) perovskites in HS (**Figs. 4-6**).[6-8] 3) The LS broad band peaking at 2.5 eV (**Fig. 4**) coincides with the centroid of the spin-allowed crystal-field bands at the spin crossover transition (S= 2→1), which corresponds to a crystal-field splitting 10*Dq* of 27*B* (*C/B* = 4.6), where *B* and *C* are the Racah parameters for $(MnF_6)^{3-}$.[6,12,13] On the other hand, there is an argument given in Ref. 1 to support a structural transition at 37 GPa instead of a HS→LS crossover transition, which is not properly justified. The authors state that the 10*Dq* required to stabilize the LS state should be higher than 3 eV, without any further justification. The value of 10*Dq* derived from the TS diagram at the transition point requires accurate knowledge of the Racah parameter *B*, which is known to be *B* = 0.097 eV (780 $cm^{-1}$) in $(MnF_6)^{3-}$ at ambient conditions.[9] This value can also be obtained from the spin-flip transition energies, since the first observed spin-flip transition depends on pressure as $E_{SP2}$ = 2.397 – 0.0018 *P* (in eV and GPa units, respectively).[6,8] From the TS diagram (**Fig. 4**)[12,13] its energy varies as 24.6 *B* in HS up to the spin crossover point. This means that *B* slightly decreases from 0.097 eV at ambient pressure to about 0.094 eV at 37 GPa, which is considered to be the spin crossover pressure as stated elsewhere.[6] This means that 10*Dq* should be about 2.54 eV –lower than 3 eV–, and considering that the single-electron $t_{2g}^4\, e_g^0 \rightarrow t_{2g}^3\, e_g^1$ transition actually gives rise to several spin-allowed transitions within the $d^4$ electronic configuration, from the LS $^3T_1$ ground state to the $^3\Gamma_i$ excited states, whose energies spread over the 0.4 eV range: 2.39 eV ($^3T_1$), 2.48 eV ($^3T_2$), 2.65 eV ($^3A_2$), and 2.76 eV ($^3A_2$).[12,13] All these transitions give rise to a broad band whose centroid peaks at about 2.54 eV, in agreement with the observed spectra of $CsMnF_4$ above 37 GPa in LS (**Fig. 4**).[6] 4) The argument given to rule out spin transition about the enthalpy difference between HS and LS states in tetragonal symmetry is not supported by any information or evidence even in the *Supporting Information*.[1] It is well known that the estimation of a spin crossover pressure from *ab initio* calculations is subtle and delicate since it requires a good knowledge of the actual HS and LS structures and the calculations are very sensitive to parameters such as the Hubbard U among others.[14,15] It is therefore scientifically irrelevant to state that the enthalpies for LS and HS are a few tenths of eV, based on an inadequate HS structure and an unknown LS structure, without providing details of the performed calculations. However, optical spectroscopy offers a more suitable means of identifying spin crossover phenomena. This technique can effectively distinguish between the distinct spectra associated with each spin state, irrespective of the specific crystal structures involved in the spin transition. This has been well established for $Fe^{2+}$ compounds and extensively documented elsewhere.[16-18]

**6) The Jahn-Teller Effect: Theorem, Theory, and Molecular Distortion.**

Finally, some concepts about the JT effect need to be clarified in order to avoid misunderstandings among scientists working with systems that exhibit JT distortions. This follows from sentences (*i-iii*) on page 13234 of Ref. 1, which state: (*i*) "The existence of a JT effect requires a degenerate electronic state in the *initial* geometry", (*ii*) "$CsMnF_4$ is a layered compound where layers are perpendicular to the crystal *c* axis (Figure 1). Accordingly, one would expect that the axis of the $MnF_6^{3-}$ unit perpendicular to the layer plane plays a singular role, a fact seemingly not consistent with the JT assumption.", and (*iii*) "The local equilibrium geometry for $MnF_6^{3-}$ in $CsMnF_4$ is not tetragonal. Indeed, even assuming Y as the main axis (Figure 1) the symmetry would be at most orthorhombic because $R_X - R_Z$ = 0.037 Å. Accordingly, one should expect four and not only three *d*−*d* transitions with $\Delta S = 0$ for $CsMnF_4$."

Because of these three sentences, it is convenient to distinguish between the JT theorem, the JT theory, and the JT distortion when dealing with the JT effect. The JT theorem states that a nonlinear polyatomic molecule with an orbital degenerate ground state is unstable and must distort to a lower energy configuration, or literally "All nonlinear configurations are therefore unstable for an orbital degenerate electronic state".[19] Based on the JT theorem, the JT theory goes further and explores the distortions of the molecule under different structural perturbations. In particular, for highly distorted $E \otimes e$ JT systems such as octahedral $MX_6$ complexes with *M*: $Cr^{2+}$, $Cu^{2+}$, $Mn^{3+}$, and *X*: $Cl^-$, $F^-$, most of which exhibit elongated octahedral distortions,[20-24] the JT theory shows that the equilibrium geometry caused by a tetragonal symmetry perturbation of the octahedron corresponds to the JT distorted geometries of the octahedron (three degenerate minima associated with tetragonally elongated octahedra along the x, y and z axes) slightly modified by the axial perturbation (**Fig. 8**).[20-24] The JT theory shows that the equilibrium coordination geometry of the JT ion varies from elongated tetragonal along the tensile perturbation direction, to two equivalent minima corresponding to the JT elongated octahedra with a slight orthorhombic distortion for a compressive $D_{4h}$ perturbation of the site, to a tetragonally compressed geometry if the compressive perturbation is large enough (**Fig. 8**). In general, a compressive $D_{4h}$ perturbation results in two equivalent elongated octahedra associated with the two JT distorted geometries with the long axes perpendicular to the $D_{4h}$ axis of the compressed site. The two short *M-X* bonds are generally different; their difference, which is zero for an $O_h$ site (three equivalent $D_{4h}$ elongated geometries), increases with the $D_{4h}$ compressive perturbation of the site, yielding local JT distortions of $D_{2h}$ symmetry. Thus, E⊗e JT theory shows that slight tetragonal distortions of the initial octahedral symmetry lead to equilibrium geometries corresponding to the JT distortions obtained in $O_h$ slightly perturbed by the $D_{4h}$ compressive strain of the site. Therefore, JT theory shows that it is possible to have equilibrium geometries of the $MX_6$ complex corresponding to JT distortions when the JT ion is placed in a $D_{4h}$ compressed –near $O_h$– symmetry, resulting in two out of three JT equivalent $D_{2h}$ elongated equilibrium geometries with their elongated *M-X* bonds perpendicular to each other and both perpendicular to the $D_{4h}$ axis of the compressive perturbation with an initial non-degenerate ground state.[21-24]

This implies a reduction in local symmetry relative to the $D_{4h}$ symmetry of the site, although the average of the two orthorhombic local symmetries retains the $D_{4h}$ symmetry of the site (**Fig. 8**). This result contradicts the authors' assertion that the local symmetry of $Mn^{3+}$ or $Cu^{2+}$ in a compressed $D_{4h}$ symmetry must remain unchanged, contrary to the JT theory.

Other structural perturbations such as non-centrosymmetric strains may be present, but they have no effect on the JT distortion since such distortion modes are not coupled to the parent-octahedral degenerate electronic $e_g$ levels in first-order perturbation.

In conclusion, JT theory shows that JT ions placed in $O_h$-perturbed low-symmetry sites, where electronic degeneracy is left, exhibit JT distortions different from the initial site symmetry in which it is placed, making the exclusion of such distortions as not being JT distortions doubtful. The authors conclude that reasons for not considering the formation of low-symmetry complexes in $Cu^{2+}$ and $Mn^{3+}$ as a JT effect are that "They are also behind the so-called plasticity property of compounds of $Cu^{2+}$ and $Mn^{3+}$ ions", citing reference 84 (Reference 25 herein). In this context, the term plasticity appears to be a euphemism, as Ref. 25, Chapter E is entitled "The Jahn-Teller and pseudo-Jahn-Teller origin of the plasticity of $Cu^{II}$ coordination sphere and distortion isomerism".

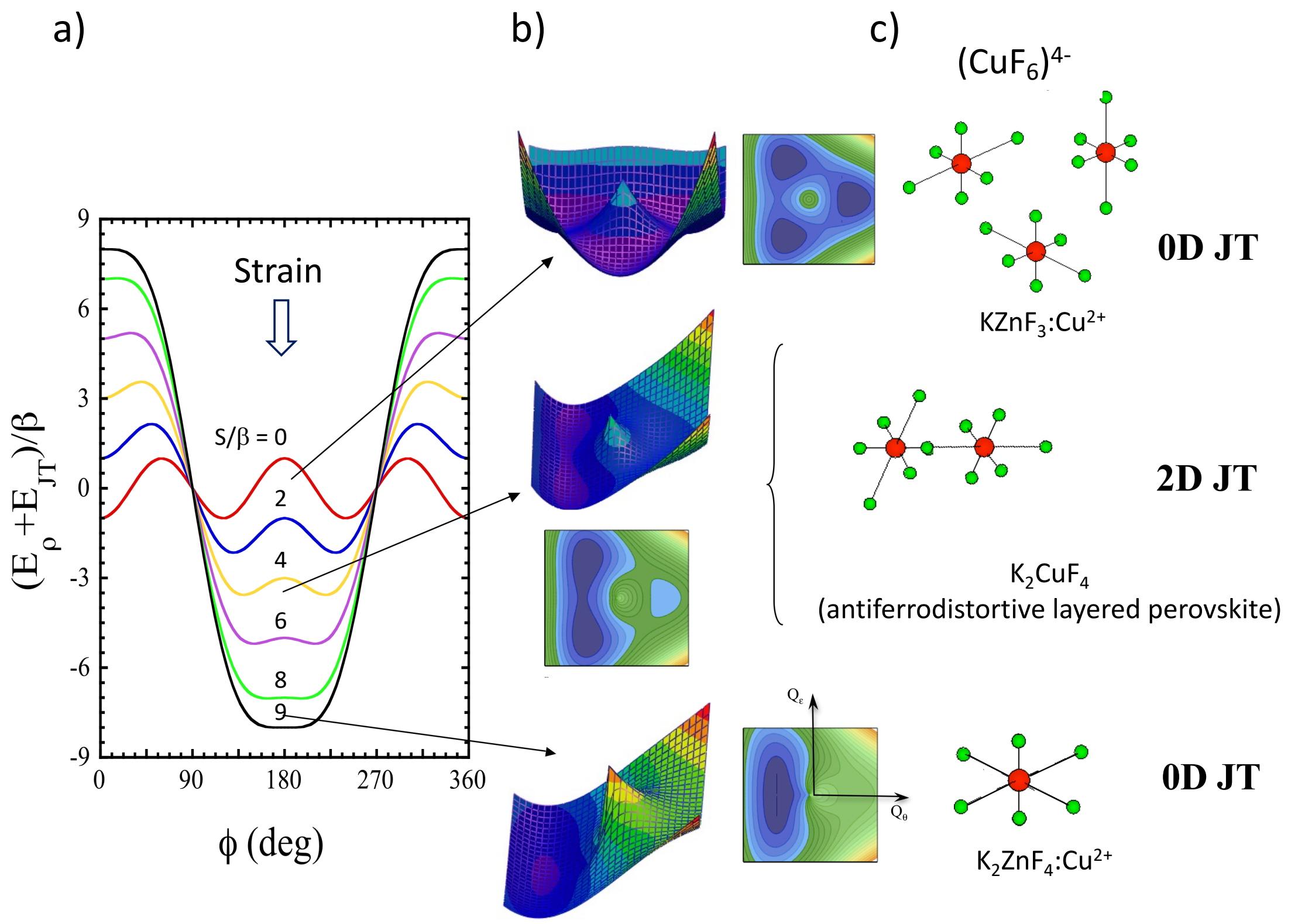

**Figure 8. Jahn-Teller and crystal anisotropy (strain) effects on the local structure of $MX_6$ (*B*,*M*: $Cu^{2+}$, $Mn^{3+}$; *X*: $F^-$, $Cl^-$) in $ABX_3$: *M* perovskite-type structures**

(**a**) Effect of crystal anisotropy induced by axial strain, characterized by the parameter $Q_{\theta S} = \rho_S \cos \phi_S$ ; $Q_{\varepsilon S} = \rho_S \sin \phi_S$ (tensile for $\phi = 0$ and compressive for $\phi = 180°$ along the $Q_\theta$-axis), on the local structure of a $(CuF_6)^{4-}$ or $(MnF_6)^{3-}$ Jahn-Teller (JT) complex in a cubic crystal. The curves represent the ground-state energy in ($Q_\theta$,$Q_\varepsilon$)-space obtained from the $E\otimes e$ JT theory. $Q_\theta$ and $Q_\varepsilon$ are the octahedral normal coordinates ($Q_\theta = \rho \cos \phi$; $Q_\varepsilon = \rho \sin \phi$), representing tetragonal and rhombic distortions, respectively. The three minima for $\rho_S = 0$ ($O_h$) correspond to three locally elongated complexes of $D_{4h}$ symmetry, with the axial distortions along *x, y,* and *z* ($\phi$ = 0°, 120°, 240°) and an equilibrium geometry given by $\rho_0$. In this model, the parameter $\beta$ (> 0 for elongated geometry minima) incorporates anharmonic and second-order JT interactions, resulting in warping of the Mexican-hat-type energy surface in ($Q_\theta$,$Q_\varepsilon$)-space. $2\beta$ is related to the energy barrier for transitions among energy minima in $O_h$ ($\rho_S = 0$). Increasing compressive strain at the JT-ion site ($\rho_S \ll \rho_0$) destabilizes the elongated complex along the *z*-axis, while the complexes elongated along $x$ and $y$ axes become topologically degenerate equilibrium geometries of $D_{2h}$ symmetry, near $D_{4h}$, with distortion $\rho_0$. The larger the tetragonal axial strain of the site ($\rho_S$), the greater the local rhombic distortion ($Q_\varepsilon$). The two degenerate wells collapse into a compressed geometry when $E_{crit} = A_e\rho_S > 9\beta$. The strain energy is introduced in the JT model as $E_{crit} = \pm A_e\rho_S$, where $A_e$ and $\rho_S$ are the electron-lattice coupling constant associated with the $E\otimes e$ JT effect and the low-symmetry coordinate at the host site.[11] The plus or minus signs represent the tensile or compressive strain energy associated with the tetragonal distortion of the site, respectively. The $\phi$–dependence of the JT energy at the equilibrium geometry ($\rho_0$) is given by:[20-23]

$$E_\rho(\phi) = -E_{JT} - \beta\cos 3\phi - S\cos(\phi - \phi_S)$$

Where $E_{JT} = -\frac{A_e\rho_0}{2}$; $\rho_0 \approx -\frac{A_e}{k}$; $\beta = A_{JT}\rho_0^2 + A_{Anh}\rho_0^3$; $S = A_e\rho_S\cos(\phi_S)$, $k$ is the force constant of the coupled vibration ($E_g$), $A_{JT}$ the second-order JT interaction, and $A_{Anh}$ (< 0) the anharmonic term. The effect of strain on the energy minima are calculated for various $S/\beta$ ratios. Note that a local $D_{4h}$ symmetry is attained for $\phi$ = 0°, 120°, 240° ($\rho_S = 0; S = 0$) corresponding to elongated distortions, or $S > 9\beta$ (compressed distortions). Any other intermediate geometry ($0 < S < 9\beta$) corresponds to local orthorhombic $D_{2h}$ symmetry.

(**b**) Ground-state energy surface E($\rho$, $\emptyset$) in ($Q_\theta$,$Q_\varepsilon$)-space for various crystal anisotropies ($S/\beta$) of $D_{4h}$ symmetry. Observe the transition from elongated to compressed coordination geometry as the axial compressive distortion of the site increases.

(**c**) Schematic views of the $(CuF_6)^{4-}$ equilibrium local structures predicted by the JT model. These predictions replicate the experimentally observed local geometries of $(CuF_6)^{4-}$ in different fluoride lattices.

Finally, it is crucial to emphasize that translating the behavior of a complex like $(MnF_6)^{3-}$ into any fluoride lattice, such as $CsMnF_4$, is plausible. This is because the phenomenon is inherently local, and the JT model parameters can effectively mimic the real structural conditions of the complex within the lattice, including crystal anisotropy, cooperative effects, and electron-lattice coupling. This approach, often referred to as cluster model, significantly simplifies the problem and provides a general framework for understanding distortions induced by the JT or pseudoJT effect.[26]